\documentclass[aps,11pt,prd,groupedaddress,nofootinbib,notitlepage,eqsecnum,preprintnumbers]{revtex4-2}
\usepackage[utf8]{inputenc}
\usepackage{hyperref}
\usepackage{xcolor}
\usepackage{graphicx}
\usepackage{amsmath,amssymb}
\usepackage{bm}
\usepackage{comment}
\usepackage[shortlabels]{enumitem}
\usepackage{overpic} 
\usepackage{ulem}
\usepackage{makecell}
\usepackage{booktabs}
\usepackage{subcaption}
\usepackage{tabularx}

\begin{document}

\title{
Probing Primordial Black Holes with upcoming Radio Telescopes: \texorpdfstring{\\ 
a case study for LOFAR2.0, FAST Core Array and BINGO}{}}

\author{Joao R. L. Santos$^{1,2,3,4}$}
\email{joaorafael@df.ufcg.edu.br}

\author{Guillem Domènech$^{3,5}$}
\email{guillem.domenech@itp.uni-hannover.de}

\author{Amilcar R. Queiroz$^{1}$}
\email{amilcarq@df.ufcg.edu.br}

\affiliation{$^{1}$Unidade Acadêmica de F\'{\i}sica, Universidade Federal de Campina Grande,\\ Caixa Postal 10071, 58429-900, Campina Grande, Para\'{\i}ba, Brazil.}

\affiliation{$^{2}$Unidade Acad\^emica de Matem\'atica, Universidade Federal de Campina Grande,\\ 58429-970,  Campina Grande, Para\'{\i}ba, Brazil.}

\affiliation{$^{3}$Institute for Theoretical Physics, Leibniz University Hannover, Appelstraße 2, 30167 Hannover, Germany.}

\affiliation{$^{4}$Institut f\"ur Theoretische Physik, Ruprecht-Karls-Universit\"at Heidelberg, Philosophenweg 16, D-69120 Heidelberg, Germany}

\affiliation{$^{5}$Max Planck Institute for Gravitational Physics, Albert Einstein Institute, 30167 Hannover, Germany.}

\begin{abstract}
Fast Radio Bursts (FRBs) are among the most intriguing phenomena observed in radio astronomy. So far, about 130 FRB signals have been confirmed and characterized by different surveys, and the CHIME telescope has recently reported a new catalog of 4539 bursts. Therefore, these numbers are expected to increase in the coming years. The detection, or lack thereof, of lensed FRB events can be used to probe Primordial Black Holes (PBHs) as a fraction of dark matter. We investigate the potential of three upcoming radio telescopes, LOFAR2.0, FAST Core Array, and BINGO, to test the PBH scenario.  We forecast that LOFAR2.0 will constrain $f_{\mathrm{PBH}} < 0.16$ for PBH masses $M_{\rm PBH}>1\,{M_{\odot}}$, while FAST Core Array and BINGO will restrict $f_{\mathrm{PBH}} < 0.39$ for $M_{\rm PBH}>10\,{M_{\odot}}$ and $M_{\rm PBH}>10^{-1}\,{M_{\odot}}$, respectively. Despite the existence of stricter constraints, FRB lensing offers an independent and complementary probe of PBHs in the Universe, which will improve in the future.

\end{abstract}

\maketitle

\section{Introduction}

FRB observations have attracted significant attention in the astronomy and physics communities since the first reported detection in 2007 of a bright, unidentified radio signal in the $1.4\,\mathrm{GHz}$ survey of the Magellanic Cloud conducted by the Parkes radio telescope \cite{Lorimer:2007qn}. After that, around a thousand possible candidates have been reported, and, at the time of writing,\footnote{Data obtained from the Transient Name Server: \url{https://www.wis-tns.org/}} 131 FRBs have been confirmed by multiple surveys \cite{jia2025} and a recent catalog with 4539 bursts was reported by CHIME (Canadian Hydrogen Intensity Mapping Experiment) telescope \cite{TheCHIMEFRB:2026nji}.  

The flux densities and luminosities of the FRBs are high, varying between $0.5$ - $100\, \mathrm{Jy}$, and 
$\sim 10^{36} \,-\, 10^{44}$ $\mathrm{erg \,s}^{-1}$ \cite{Curtin_2024}, and of short time durations, usually around $\mathrm{ms}$ - $\mathrm{\mu s}$. The short time window poses an experimental challenge, namely how to determine an optimal backend resolution for measuring these bursts. The events are mainly extragalactic and usually unique. The fact that we are observing mainly single bursts makes it difficult to understand the origin of such phenomena. 

The leading explanation is that FRBs come from neutron stars, specially magnetars \cite{curtin2025, Cordes:2015fua, Popov2007,Pen:2015ema}, although there are alternative explanations involving Primordial Black Holes (PBHs) swallowing neutron stars \cite{Abramowicz:2017zbp,Fuller:2017uyd,Kainulainen:2021rbg,Amaral:2023ekd,Carr:2023tpt} and the merger of charged PBHs \cite{Deng:2018wmy,Meng:2024cwa}. Moreover, another interesting question is whether repeating and non-repeating FRBs represent distinct classes of objects \cite{curtin2025}.

The time delay of the FRB signal across different radio frequency channels depends on the interaction of free electrons along the signal's path. Higher-wavelength fronts arrive later than shorter ones. This effect is parametrized by the so-called dispersion measure (DM). From the DM, one can estimate the FRB's redshift and characterize the matter along the line of sight. As most of the detected FRBs are extragalactic in origin, they may also be used to constrain cosmological models \cite{Kalita:2024xae}, estimate the fraction of baryonic mass in the intergalactic medium (IGM) \cite{Munoz:2018mll,Walters:2019cie,Qiang:2020vta}, and to measure the Hubble parameter \cite{Kalita:2024xae}.

Observations of FBRs by the new generation of radio telescopes may be a promising path for investigating and characterizing the abundance of compact dark matter objects, like PBHs, as pioneered by Mu\~{n}oz et al. \cite{Munoz_2016}. Since then, several investigations have been published in the literature over the last few years, as we can see, for instance, in \cite{Brandt:2016aco, Wang:2018ydd, Xiao:2022hkl, Leung_2022, Kalita_2023, Profumo_2025}. In particular, the CHIME Collaboration constrained the fraction of PBHs as dark matter to $f_{\rm PBH}<0.8$ for $M\sim 10^{-3}\, M_\odot$ \cite{Leung_2022}. Moreover, a recent work by Huan Zhou et al. \cite{Zhou:2026flq} unveiled evidence for intermediate-mass PBHs through microlensing effects on two FRBs presented in the new CHIME catalogue. The inferred lens masses derived from these FRBs are between $539-609\,M_{\odot}$ and $1544-2571\,M_{\odot}$ for a fraction ratio $\sim 4\,\%$. These results reveal a tension with several constraints, such as those from LIGO/VIRGO/KAGRA \cite{Green:2020jor, Oguri:2022fir}, OGLE \cite{Mroz:2024wag}, and Cosmic Microwave Background \cite{Serpico:2020ehh}. For studies investigating the impact of modifications of gravity, and different mass functions and FRB distributions see Refs.~\cite{Kalita_2023} and \cite{Li_2025}, respectively.

PBHs may form by the collapse of large fluctuations in the primordial plasma, as first pointed out by Zel’dovich and Novikov \cite{Zeldovich:1967lct} and Hawking and Carr \cite{Hawking:1971ei,Carr:1974nx}. Depending on their mass, PBHs may explain all the dark matter \cite{Carr:2020xqk, Khlopov:2008qy, Belotsky:2014kca}, the OGLE and HSC microlensing candidate events \cite{Mroz:2017mvf,Niikura:2019kqi,subaru_2026}, some of the LIGO/Virgo/KAGRA GW events \cite{Hutsi:2020sol,Berti:2025usa, Belotsky:2018wph} and the seeds of supermassive black holes \cite{Kawasaki:2012kn,Bernal:2017nec,Byrnes:2024vjt}. For recent reviews on PBHs see Refs.~\cite{Sasaki:2018dmp,Green:2020jor,Carr:2023tpt,Casanueva-Villarreal:2025kmd,Byrnes:2025tji}. In the mass range of our interest, that is $M>10^{-3} M_\odot$, the PBH fraction is currently constrained by lensing of SN type Ia \cite{Zumalacarregui:2017qqd}, microlensing of stars in the Magellanic Clouds \cite{Mroz:2024mse,Mroz:2025xbl}, gravitational lensing for dressed PBHs \cite{Oguri:2022fir}, caustic crossings \cite{Oguri:2017ock}, LVK binary black hole mergers \cite{Andres-Carcasona:2024wqk} and the Cosmic Microwave Background (CMB) \cite{Serpico:2020ehh,Agius:2024ecw}. Current constraints range from $f_{\rm PBH}<10^{-2}$ for sub-solar mass PBHs down to $f_{\rm PBH}<10^{-8}$ for intermediate mass PBHs. For a detailed summary of constraints see Refs.~\cite{Carr:2020gox,Profumo_2025,Carr:2026hot}. Nevertheless, it is important to stress that FRB lensing offers an independent and complementary way to probe PBHs in these mass ranges, and that these test will improve in the future.

In this work, we apply the general formalism of previous works to provide a quantitative, focused study of the potential to probe PBHs with the following upcoming radiotelescopes:  LOFAR (LOw Frequency ARray) and FAST (Five-hundred-meter Aperture Spherical radio Telescope), which are undergoing upgrades, and BINGO (Baryon Acoustic Oscillations from Integrated Neutron Gas Observations), which is under construction in Brazil. The upgrade to LOFAR, called LOFAR2.0 \cite{lofarwhitepaper,lofar20}, aims to double its observing speed and expand its angular resolution \cite{Lofar_2026}. The FAST Core Array upgrade will image FRBs and pulsars with much greater detail, enabling better characterization and revealing their origins and evolution \cite{FAST-Core-Array}. These upgrades are expected to be completed by 2026 and 2030, respectively.

BINGO will be one of the first fixed transit telescopes dedicated to exploring BAO at this frequency band \cite{Abdalla_2022}. In addition to the main radio telescope, there will be a series of auxiliary radio telescopes, forming an interferometric array that will not only enhance the quality of BAO data measurements but also enable BINGO to become an excellent machine for measuring transient radio phenomena such as pulsars and FRBs \cite{dosSantos:2023ipw}. BINGO is in its final stages of construction, and it is expected to start operating in the coming years. BINGO is also planning a joint operation together with FAST, Tianlai, ASTRON, and Nan\c{c}ay Observatory, in a phase called BINGO/ABDUS (Advanced Bingo Dark Universe Studies) \cite{bingoabdus-2023}. The BINGO/ABDUS collaboration intends to increase the redshift range of BINGO, observing radio signals from $z\sim 2.1$, and covering a larger fraction of the sky in a resolution between $27^\prime - 40^\prime$ \cite{bingoabdus-2023}. Thus, LOFAR, FAST, and BINGO will increase the number of FRB detections in the coming years and support the characterization of FRBs detected by other surveys, such as CHIME. 
 
To derive the forecasts, we apply the work of Mu\~{n}oz et al. \cite{Munoz_2016} and study the impact of the signal-to-noise (SNR) ratio, the total number of measured bursts, and the time resolution of these different radio telescopes in the computation of the optical depth and of the fraction of PBHs. These specifications of the radio telescopes were implemented following the steps introduced by Leung et al. \cite{Leung_2022} and Kalita et al. \cite{Kalita_2023}. Furthermore, our forecast is based on the most recent catalog of 131 FRBs recently provided by Jia et al. \cite{jia2025} and detailed in App.~\ref{append_a}.

This work is organized as follows. In Section \ref{sec3}, we review  generalities on FRB lensing as probes of PBHs. In section \ref{sec4}, we present the details about each radio telescope, our forecast for the optical depth and for the fraction of PBHs, and a summary of our results. Section \ref{final_remarks} is dedicated to our conclusions and discussion.  Throughout this work, we use the following constants and parameters in our numerical calculations: $c = 299{,}792{,}458\,\mathrm{m/s}$, $G = 6.67430\times10^{\,-11}\,\mathrm{m^3\,kg^{-1}\,s^{-2}}$, $M_{\odot} = 1.98847\times10^{30}\,\mathrm{kg}$, $H_0 = 67.7\,\mathrm{km\,s^{-1}\,Mpc^{-1}}$. When needed, we use the cosmological parameters from the best fit of Planck 2018 \cite{aghanim2020planck}, that is $\Omega_M = 0.30966$, $\Omega_\Lambda = 0.68884$, $\Omega_b=0.04897$, and $\Omega_c = 0.26069$. $\Omega_M$, $\Omega_b$, $\Omega_c$ and $\Omega_\Lambda$ are the density fraction of matter, baryons, dark matter and dark energy in the Universe, respectively.

\section{Review on FRBs Lensing} 
\label{sec3}

In this section we present the generalities of FRBs characterization, gravitational lensing by a point source and the optical depth of a survey.

\subsection{FRBs characterization}
\label{sec2}

One of the most challenging parameters to identify using FRBs' data is the redshift of each source, which can be estimated from the DM. The calculation of $\rm DM$ typically involves various contributions, primarily due to the host galaxy, the intergalactic medium, and the Milky Way's interstellar medium. The observed $\rm DM$ is computed by \cite{Petroff:2014taa}
\begin{equation}
{\rm DM}_{\mathrm{obs}}(z)= {\rm DM}_{\mathrm{MW}}+{\rm DM}_{\mathrm{IGM}}+\frac{{\rm DM}_{\textrm{host}}}{1+z}\;,
\label{e1}
\end{equation}
where ${\rm DM}_{\rm MW} = {\rm DM}_{\rm MW,ISM}+{\rm DM}_{\rm MW,halo}$ is the dispersion measure of the Milky Way, with $\rm DM_{\rm{MW,ISM}}$ as the galactic interstellar medium $(\sim 10^0-10^3\, {\rm pc\ cm^{-3}})$ and $\rm DM_{\rm{MW,halo}}$ the galactic halo contributions, respectively. The Milky Way halo contribution DM$_{\textrm{MW,halo}}$ is constrained by different surveys as lying between  $50-117\, {\rm pc\ cm^{-3}}$ \cite{James_2021, Leung_2022}. Furthermore, the dispersion measure related to the intergalactic medium, $\rm DM_{\rm{IGM}}$ varies between  $(\sim 10^2-10^3\, {\rm pc\ cm^{-3}})$, while the one corresponding to the host galaxy $\rm DM_{\rm{host}}$ covers the range $(\sim 10^0-10^3\, {\rm pc\ cm^{-3}})$. Note that the factor $(1+z)$ in \eqref{e1} is due to cosmic expansion.

The dispersion measure $\rm DM_{\rm IGM}$ is determined using the Macquart relation \cite{Macquart:2020lln}, whose explicit form is

\begin{equation} 
 {\rm DM}_{\rm IGM}(z) = A \Omega_b H_0^2 \int_0^z \frac{\left(1+z\right) x_e\left(z\right)}{H_0\,\sqrt{\Omega_M (1+z)^{3}+\Omega_\Lambda}} d z\,; \qquad A=\frac{3cf_{\rm IGM}}{8 \pi G m_p}\,,
\label{e2}
\end{equation}
where $f_{\rm IGM}=0.83$ is the baryon mass fraction in the IGM \cite{shull2012baryon}, $H_0$ is the Hubble parameter today, and $m_p$ is the proton mass. Lastly, in Eq.~\eqref{e2}, $x_e(z)$ denotes the free electron fraction (or degree of plasma ionization), which is given by
\begin{equation}
  x_{\rm e}(z)=Y_{\rm H} x_{\rm e, H}(z)+\frac{1}{2}Y_{\rm He} x_{\rm e, He}(z)\;, 
  \label{e3}
\end{equation}
where $Y_{\mathrm{H}}=3/4$, $Y_{\mathrm{He}}=1/4$, $x_{\mathrm{e}, \mathrm{H}}(z)$ and $x_{\mathrm{e}, \mathrm{He}}(z)$ are the mass fractions and the ionization fractions of hydrogen and helium, respectively. In our analysis, we consider $x_{\rm e, H}(z)=x_{\rm e, He}(z)=1$ since hydrogen and helium are fully ionized at $z<3$ \cite{Meiksin:2007rz,becker2011detection},  yielding $x_{\rm e}=7/8$. The previous relations are used to estimate the redshift of the source for different FRB pulses, which are presented together with the values of $\rm DM_{obs}$ and ${\rm DM}_{\rm MW} $ in App.~\ref{append_a}.

\subsection{Gravitational lensing by a point source}

The procedures presented in this section and in the computation of the total optical depth and of the $f_{\mathrm{PBH}}$ are based on the works of Mu\~noz et al. \cite{Munoz_2016}, Leung et al. \cite{Leung_2022}, and Kalita et al. \cite{Kalita_2023}. To characterize lensing effects due to PBHs, we treat of them as a point lens, with the standard angular Einstein radius given by \cite{Bartelmann:2010fz,Munoz_2016}
\begin{equation}
    \theta_E=2\,\sqrt{\frac{G\,M_L}{c^2}\,\frac{D_{LS}}{D_S\,D_L}}\,,
\end{equation}
where $D_S$, $D_{L}$ and $D_{LS}$ are the distances to the source, to the lens and between the source and the lens, respectively. The deflection angle, the normalized impact parameter $y$, and the angular impact parameter $\beta$ are defined as
\begin{equation} \label{s3_eq05}
   \alpha = \frac{\theta_E^2}{\theta}\,; \qquad y = \frac{\beta}{\theta_E}\,; \qquad \beta = \theta - \frac{\theta_E^2}{\theta}\,.
\end{equation}
A point lens will produce two images located at
\begin{equation} \label{s3_eq1}
    \theta_{\pm} = \left(\frac{\beta}{2} \pm \frac{1}{2}\sqrt{\beta^2+4\theta_E^2}\right)\,.
\end{equation}

For the magnification ratio we adopt the approach of Kalita et al. \cite{Kalita_2023}, which yields
\begin{equation}
    \mu = \left(\frac{y+\sqrt{y^2+4}}{y-\sqrt{y^2+4}}\right)^2\,,
\end{equation}
enabling us to constrain $y_{\mathrm{max}}$ as
\begin{equation} \label{s3_eq1_5}
    y_{\mathrm{max}} = \frac{q-1}{\sqrt{q}}\,; \qquad q =\sqrt{\mu_{\mathrm{max}}}\,.
\end{equation}
As a maximum value of the magnification ratio, we adopt $\mu_{\mathrm{max}} = 1/3 \,\mathrm{SNR}$, where $\mathrm{SNR}$ depends on the radio telescope. This maximum value was introduced by Sammons et al. \cite{Sammons:2020kyk} and applied in \cite{Leung_2022, Kalita_2023}. The factor $1/3$ corresponds to a realistic threshold for the signal-to-noise ratio at which the burst can be detected in autocorrelation \cite{Sammons:2020kyk, Leung_2022}. For such a threshold, the signal can be distinguished from noise fluctuations \cite{Kalita_2023}. Therefore, this parameter depends directly on the design of the instruments considered in this forecast. Note that Eq. \eqref{s3_eq1_5} circumvents the need to set a threshold value for the flux ratio in the computation of $y_{\mathrm{max}}$, yielding a more realistic assumption on the forecast.

\begin{figure}
    \centering
    \includegraphics[width=0.4\columnwidth]{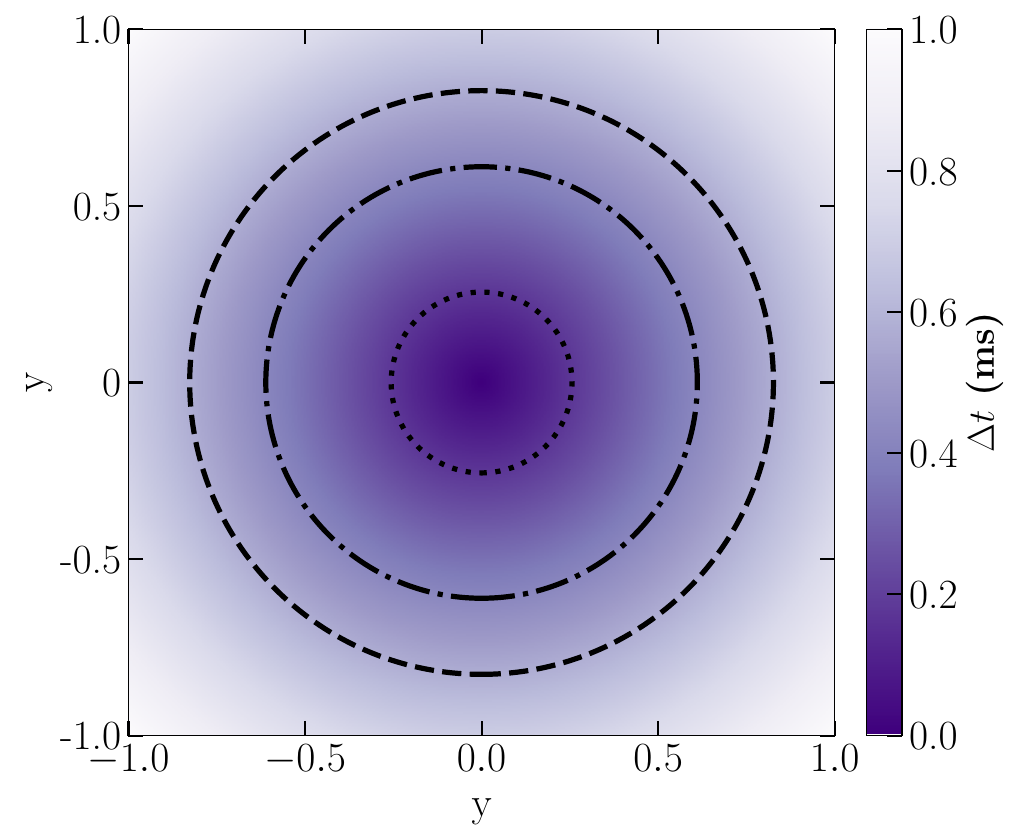}
    \caption{Lens plane to characterize the detectability of $\Delta t$ with different SNR. The figure was depicted considering $z_L=0$, $M_L = M_{\odot}$, $\mathrm{SNR} = 5$ (dotted annulus), $\mathrm{SNR} = 10$ (dashed-dotted annulus), $\mathrm{SNR} = 15$ (dashed annulus), and $\Delta t$ variations are scaled in $\mathrm{ms}$.}
    \label{fig:01}
\end{figure}

The presence of a standard Schwarzschild lens will result in a time delay between the images, given by \cite{Bartelmann:2010fz,Munoz_2016}
\begin{equation} \label{sec3_eq2}
    \Delta t = \left(1+z_L\right)\,\frac{D_\Delta\,\theta_E^2}{c}\,\left[\frac{y}{2}\,\sqrt{y^2+4}+\mathrm{ln}\left(\frac{\sqrt{y^2+4}+y}{\sqrt{y^2+4}-y}\right)\right]\,,
\end{equation}
where $z_L$ is the redshift of the lens and $D_\Delta = D_L\,D_S/D_{LS}$. We illustrate the features of the time delay in Fig.~\ref{fig:01}, considering $z_L=0$, $M_L = M_{\odot}$, $\mathrm{SNR} = 5$ (dotted annulus), $\mathrm{SNR} = 10$ (dashed-dotted annulus), $\mathrm{SNR} = 15$ (dashed annulus). $\Delta t$ variations are scaled in $\mathrm{ms}$. From Fig.~\ref{fig:01}, we see that larger SNR allow the radio telescopes to observe larger values of time delay, corresponding to an increase in the maximum values of $y$. As we will reveal in the forecast results, the SNR will be crucial for setting bounds on the maximum mass fraction of PBHs.

\subsection{Event rate and optical depth}

Let us now compute the probability of a FRB located at a redshift $z_S$ to be lensed. The optical depth of this radio transient reads \cite{Munoz_2016}
\begin{equation} \label{sec4_eq2}
    \tau(M_L,z) = \int_{0}^{z}\,d\chi(z_L)\,(1+z_L)^2\,n_L\,\sigma(M_L,z_L)\,,
\end{equation}
where 
\begin{equation}
 \chi(z) = c\,\int_0^{z}\,\frac{dz}{H(z)}\,, \qquad H(z) = H_0\sqrt{\Omega_M\,(1+z)^3+\Omega_{\Lambda}}\,,   
\end{equation}
which are the comoving distance and the Hubble function, respectively. Moreover, $n_L$ is the comoving number density of lenses and 
\begin{equation} \label{sec4_eq3}
    \sigma(M_L, z_L) = \frac{4\,\pi\,G\,M_L}{c^2}\,\frac{D_L\,D_{LS}}{D_{S}}\,\left(y_{\mathrm{max}}^2 (\mu)-y_{\mathrm{min}}^2(M_L,z_L)\right)\,,
\end{equation}
is the lensing cross-section of a point lens with mass $M_{L}$. 

For simplicity, let us assume a monochromatic mass function with a peak at $M_L$. In that case, one has that
\begin{equation} \label{sec4_eq4}
 \frac{dn_{L}}{dM} = \frac{\rho_0}{M_L}\,\,f_{\mathrm{PBH}}\,\Omega_c \,\delta (M-M_L)\,;  \qquad n_{L} = \frac{ \rho_0\,f_{\mathrm{PBH}}\,\Omega_c}{M_{L}}\,,
\end{equation}
where $f_{\mathrm{PBH}}$ is the fraction of PBHs as dark matter and $\rho_0 = 3\,H_0^2/8\pi G$ is the critical density of the Universe. By following the approach of \cite{Munoz_2016, Leung_2022, Kalita_2023}, we substitute Eqs.~\eqref{sec4_eq3} and \eqref{sec4_eq4} into Eq.~\eqref{sec4_eq2}, yielding us to rewrite the lensing optical depth as
\begin{equation}\label{sec4_eq4_5}
    \tau(M_L,z) = \frac{3}{2}\,f_{\mathrm{PBH}}\,\Omega_c\, \int_{0}^{z}\,d z_L\,\frac{H_0^2}{c\,H(z_L)}\,\frac{D_L\,D_{LS}}{D_S}\,(1+z_L)^2\,\left(y_{\mathrm{max}}^2 (\mu)-y_{\mathrm{min}}^2(M_L,z_L)\right)\,,
\end{equation}
where $y_{\mathrm{min}}$ is obtained by solving Eq. \eqref{sec3_eq2} for $\Delta t_{\mathrm{min}}$. The values used for $\Delta t_{\mathrm{min}}$ depend on the better time resolution of each radio telescope considered in this forecast.

\begin{figure}
    \centering
    \includegraphics[width=0.45\columnwidth]{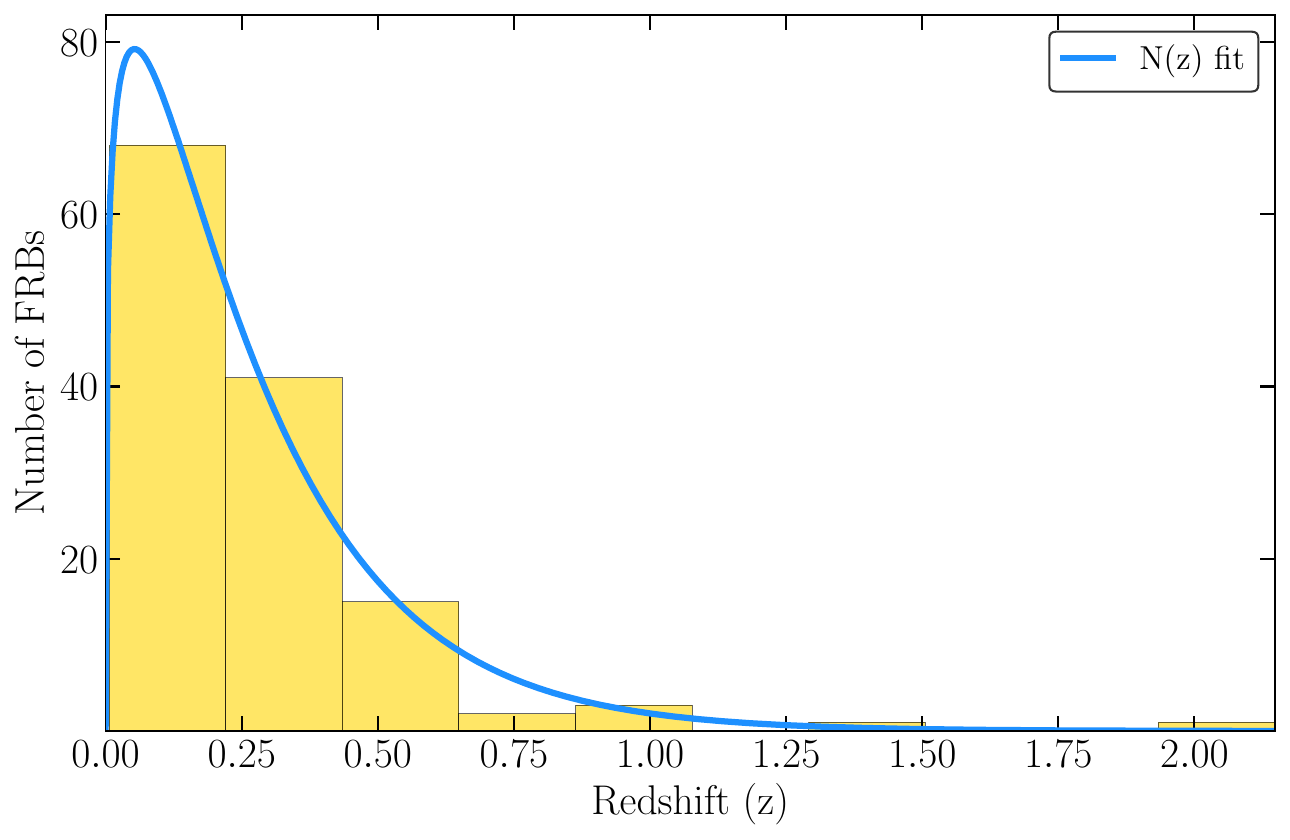}
    \caption{Histogram of the recent data of 131 confirmed FRB events. The fit \eqref{sec4_eq5} is shown with the solid blue line. The dataset used is presented in App.~\ref{append_a}.}
    \label{fig:02}
\end{figure}

The integrated optical depth is such that
\begin{equation}\label{eq:opticaldepthint}
 \bar{\tau}(M_L) = \int\,dz\,\tau(M_L,z)\,\bar{N}(z)\,,
\end{equation}
where $\bar{N}(z)$ is the normalized distribution function, which is given by
\begin{equation}
 \bar{N}(z) = \frac{N(z)}{\bar{N}_0}\,;\qquad \bar{N}_0 = \int_0^{z_{\mathrm{max}}}dz N(z)\,.  
\end{equation}
here $z_{\mathrm{max}}$ stands for the maximum redshift value of the detected FRBs, and the normalization is computed before the determination of $\bar{\tau}$.
Since the number of cataloged FRB is smaller than $10^4$, which would be the ideal amount of events to properly constrain a lensing forecast \cite{Munoz_2016}, we fit the present FRB dataset with an analytic distribution. In App.~\ref{append_a}, we present the most recent data of 131 confirmed FRB pulses, which were used to build the histogram illustrated in Fig.~\ref{fig:02}. The tables in App.~\ref{append_a} represent a combination of two datasets previously used by references \cite{Sales:2025shu} and \cite{jia2025} to constrain Hubble and cosmographic parameters.

We find that a fit to  $N(z)$ in the histogram of Fig.~\ref{fig:02} is given by a Gamma distribution, whose explicit form is
\begin{equation} \label{sec4_eq5}
    N(z) = N_0\,z^{a_1}\,e^{-a_2\,z}\,; \qquad N_0 = 208.7210\,; \qquad a_1 = 0.2465 \,; \qquad a_2 = 4.6269\,.
\end{equation}
To reach this fit, we tested three different distributions: the Lognormal, Gamma, and Weibull. The tests consisted of computing the log-likelihood of the redshift data and evaluating how well each distribution fits the data. Then, we applied the Akaike Information Criterion (AIC) to find the distribution corresponding to the minimal value of this criterion. Such a procedure enabled us to determine the non-Gaussian distribution introduced in Eq.~\eqref{sec4_eq5}. Note that, although the analytical fit may improve in the future, we do not expect our results to change substantially.

\section{Forecasts for upcoming radio telescopes} \label{sec4}

Before presenting the results of our forecast, let us briefly introduce each of the radio telescopes considered here and justify the parameters chosen as our constraints.

\begin{table}
    \centering
    \setlength{\tabcolsep}{10pt}
    \renewcommand{\arraystretch}{1.25}
    \begin{tabular}{|c|c|c|c|c|}
    \hline
     Telescope  & SNR   & Best time resolution & Max number of FRBs   & Reference\\
     \hline\hline
      LOFAR2.0 & $7$  & $5.12\,\mu {\rm s} - 983\,\mu {\rm s}$ & 4500 & \cite{van_Haarlem_2013, Chawla_2025}  \\
      \hline
      FAST Core Array & $10$  &  $270\,\mu {\rm s} -128\,{\rm ms}$  & 900 & \cite{Zhang_2023,bingoabdus-2023} \\
      \hline
      BINGO & $10$  &  $1\,\mu {\rm s}-1\,{\rm ms}$  & 900  & \cite{Abdalla_2022, Wuensche:2021dcx, dosSantos:2023ipw, bingoabdus-2023}  \\
      \hline
    \end{tabular}
    \caption{Design features of the three radio telescopes considered in the forecast, and the estimated maximum number of FRBs that each radio telescope could measure in the near future.}
    \label{tab:01}
\end{table}

\subsection{The LOFAR, FAST and BINGO radio telescopes}

We summarize the design features of each radio telescope implemented in the numerical integrations in Table \ref{tab:01}, which we use to determine the optical depth and bounds on the fraction of PBHs. The public data employed is reported in Refs.~\cite{van_Haarlem_2013, Chawla_2025,Zhang_2023,bingoabdus-2023,Abdalla_2022, Wuensche:2021dcx, dosSantos:2023ipw}. We describe below each radio telescope separately.\medskip

\textbf{i) LOFAR} is a radio telescope based in the north of the Netherlands that started its operations in 2012. The telescope consists of an array of omnidirectional antennas forming sets of stations, and was originally designed to explore a frequency range between $10 - 240\,\mathrm{MHz}$. In addition to its main site, LOFAR has 51 stations across the Netherlands, Germany, France, the United Kingdom, Ireland, and Sweden.  The main scientific motivation for its construction was to explore $21\, \mathrm{cm}$ signals from the reionization, covering $z = 6 - 20$, and from the Cosmic Dawn ($z=20 - 50$). As secondary objectives, LOFAR's design is also suitable for detecting signals from ultra-high-energy cosmic rays, conducting surveys of pulsars and radio transients, searching for high-redshift radio sources, and detecting FRBs \cite{van_Haarlem_2013}.

In respect to the detection of FRBs, LOFAR has proven to be a great machine for characterizing these transient phenomena, covering the $150\,\mathrm{MHz}$ activity frequency band of 14 repeating FRBs measured by LOFAR and CHIME radio telescopes \cite{Chawla_2025}. The data were cataloged between 2012 and 2020 from the so-called LOTAAS (LOFAR Tied-Array All-Sky Survey) \cite{Lofar_2019}, with a frequency resolution of $48.8\,\mathrm{kHz}$ and a time resolution of $0.983\,\mathrm{ms}$.

In the next few years, LOFAR will be fully upgraded to LOFAR2.0, increasing its computational capacity and resolution, allowing it to search for FRBs over hundreds of tied-array beams. After such an update, the telescope expects to detect $0.3-9$ FRBs per week with $\mathrm{SNR} = 7$, and with possible nanosecond time resolution \cite{Acharya:2025ubt}, covering the redshift interval $1<z<3$. The actual minimal time resolution of the radio telescope is $5.12\,\mathrm{\mu s}$ \cite{van_Haarlem_2013}, which is one of the constraints used in our forecast, as can be seen in Table \ref{tab:01}.\medskip  

\textbf{ii) FAST} is the largest single-dish radio telescope in operation worldwide. It was completed in 2016 and operates in a frequency range between $70\,\mathrm{MHz} - 3\,\mathrm{GHz}$. FAST also has an ultra-wideband receiver operating between $270 - 1620\,\mathrm{MHz}$ and 19 dual-polarized beams working in the frequency range of $1050 - 1450\,\mathrm{MHz}$. The main scientific objectives of the FAST telescope are to conduct a large-scale neutral hydrogen survey, observe pulsars and timing arrays, and detect interstellar molecules. 

FAST was not initially designed to detect FRBs. In fact, its relatively small field of view makes the observation of such transients challenging compared with surveys fully dedicated to FRBs, such as the CHIME radio telescope. Nevertheless, FAST has extremely high sensitivity (more than twice that of the Arecibo radio telescope), and it can be useful for characterizing pulsars and FRBs by improving the accuracy of their positioning and detecting the high-precision neutral hydrogen absorption line generated by these transients in real time. FAST detects fast radio bursts using its real-time FRB searching system, which recently was able to characterize pulses with a dispersion measure range of $20-1000\,\mathrm{pc\, cm^{-3}}$, a frequency resolution of $122.07\,\rm kHz$, SNR equals to $10$, and time resolution between $128\,{\rm ms}-270.336\,\mu {\rm s}$ \cite{Zhang_2023}. Some of these parameters are presented in Table \ref{tab:01}, and were considered in our forecast.

In the next few years, FAST will be upgraded with the construction of the so-called FAST Core Array \cite{FAST-Core-Array}. This new array will consists in 24 secondary antennas with 40 m diameter, installed in a radius of 5 km from the FAST radio telescope. This new array will increase the resolution and the sensitivity of FAST, allowing it to localize transient bursts, which may unveil new insights about the sources of FRBs.\medskip

\textbf{iii) BINGO} is a radio telescope in its final phase of construction at the city of Aguiar, located in the Northeast of Brazil. BINGO will be a single-dish telescope with a $40\,{\rm m}$ primary mirror, and an array of 28 horns (receivers). The main objective of the project is to observe the $21\, \mathrm{cm}$ line associated with the hyperfine emission of the hydrogen atom. The horns are designed to operate in an optimized range between $980 \, - \, 1260\,\mathrm{MHz}$, corresponding to $z=0.13\,-\,0.48$. The radio telescope will cover approximately 5,324 square degrees of the sky over a 5-year observation cycle \cite{Abdalla_2022, bingoabdus-2023}. BINGO will be part of a new generation of radio telescopes dedicated to investigating the dark sector in a redshift range much closer to the present. Moreover, it represents one of the most relevant experiments developed and operated in Brazil.

Beyond BAO signals, BINGO also aims to be an interesting machine for detecting and localizing FRBs. To do so, it needs to be improved with a set of auxiliary radio telescopes (outriggers) and with a complementary digital backend setup covering a time resolution between $1\mathrm{\mu s} \, - \, 1 \mathrm{ms}$. The expected SNRs vary between $5\,-\,15$ and this interferometric array forecasts the detection of approximately $70$ events per year \cite{dosSantos:2023ipw}. The number of detections can be increased to $900$ events by considering a joint collaboration between BINGO, ASTRON, Nan\c{c}ay Observatory, FAST, and Tianlai radio telescopes, in a phase named BINGO-ABDUS (Advanced BINGO Dark Universe Studies) \cite{bingoabdus-2023}. These values were used as thresholds for our forecast of the fraction of PBHs, and are shown in Table \ref{tab:01}.

\subsection{Forecasts for PBH constraints}

We use the values given in Tab.~\ref{tab:01} to forecast the expected constraints on the PBH abundance. In our numerical analysis, we set a minimum and maximum value for the PBH mass given by $M_{\mathrm{min}} = 10^{-2}\,M_{\odot}$ and $M_{\mathrm{max}} = 3\times 10^{4}\,M_{\odot}$ respectively \cite{Leung_2022}. The lower limit, $M_{\mathrm{min}}$, is set by the highest time resolution of BINGO and LOFAR of roughly $1\mu{\rm s}$. Smaller lens masses lead to a too short time delay. The high mass cutoff, $M_{\mathrm{max}}$, is a consequence of the limitation in the sensitivity of the radio telescopes \cite{Leung_2022}, and for our approach it corresponds to $\Delta t_{\mathrm{max}} \sim 0.3\,s$, as the maximum time delay between the lensed images. Moreover, for $M_{\mathrm{max}} > 10^{4}\,M_{\odot}$, there are extremely strong constraints on $f_{\rm PBH}$ from CMB spectral distortions \cite{Byrnes:2024vjt,Pritchard:2025yda}. We used $90$ and $131$ points for integrating over the mass and the redshift grids, respectively.

\begin{figure}
    \centering
    \includegraphics[width=0.329\linewidth]{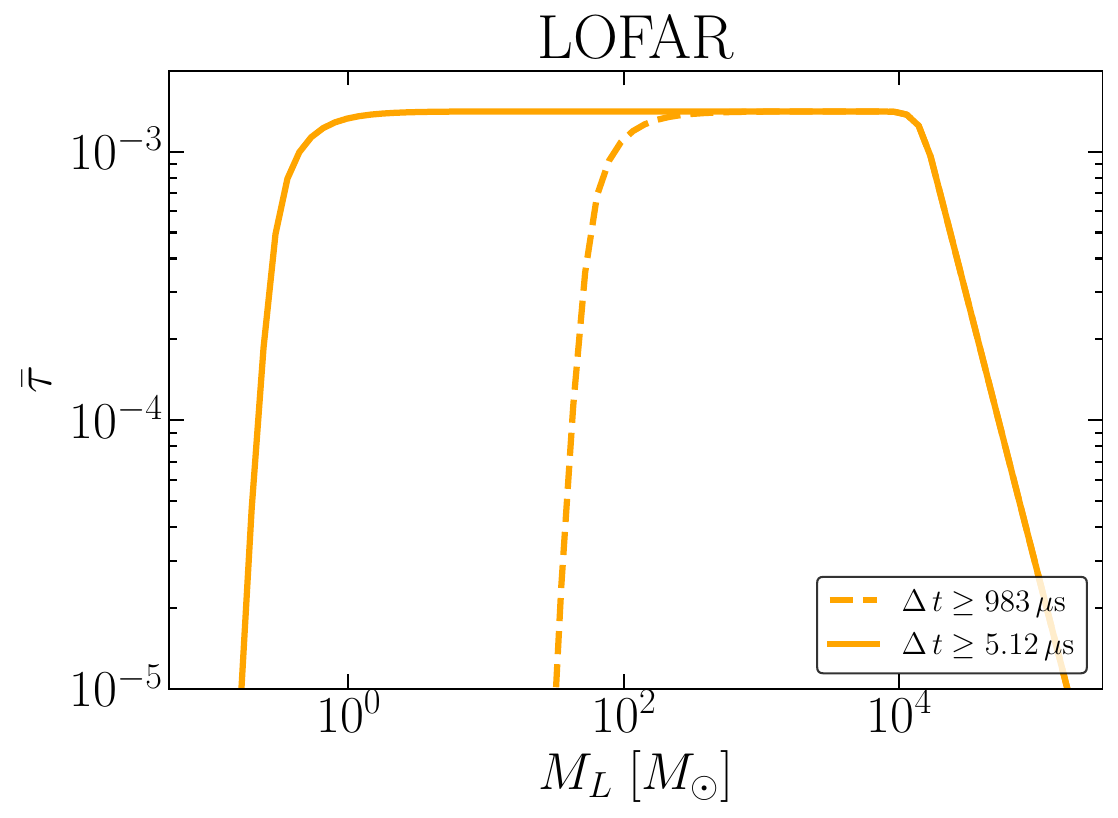} \includegraphics[width=0.329\linewidth]{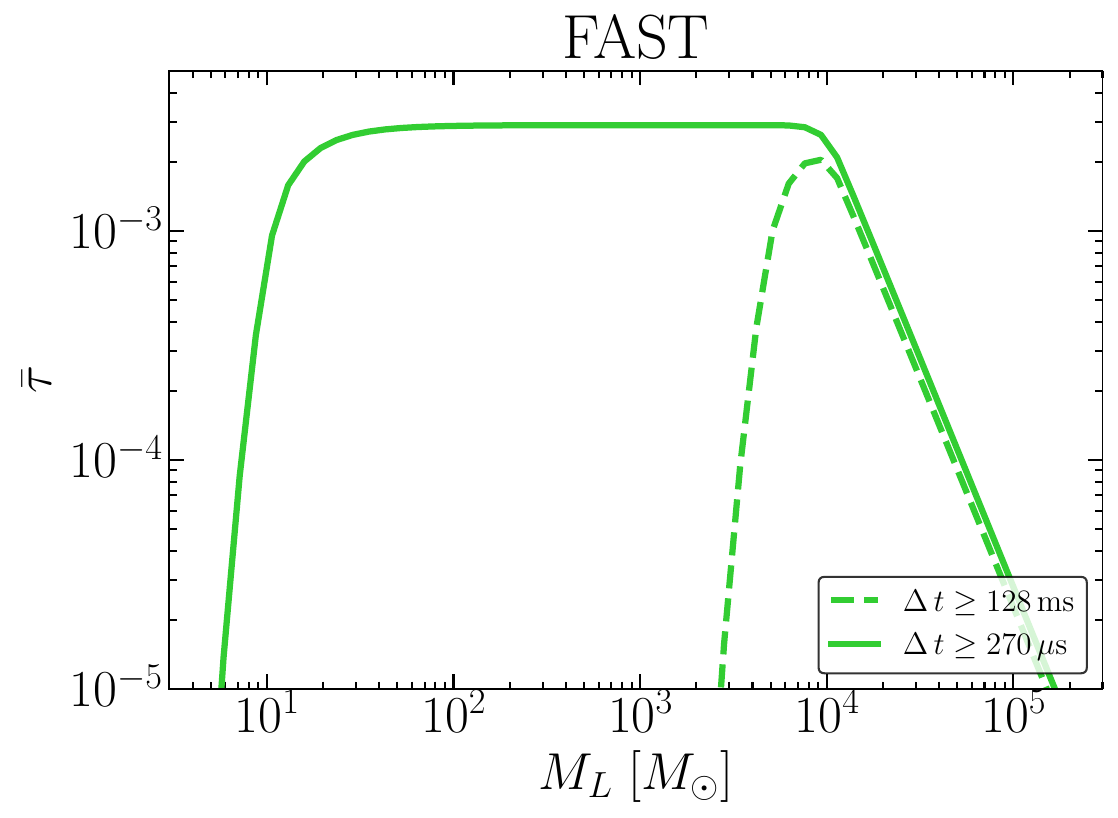} \includegraphics[width=0.329\linewidth]{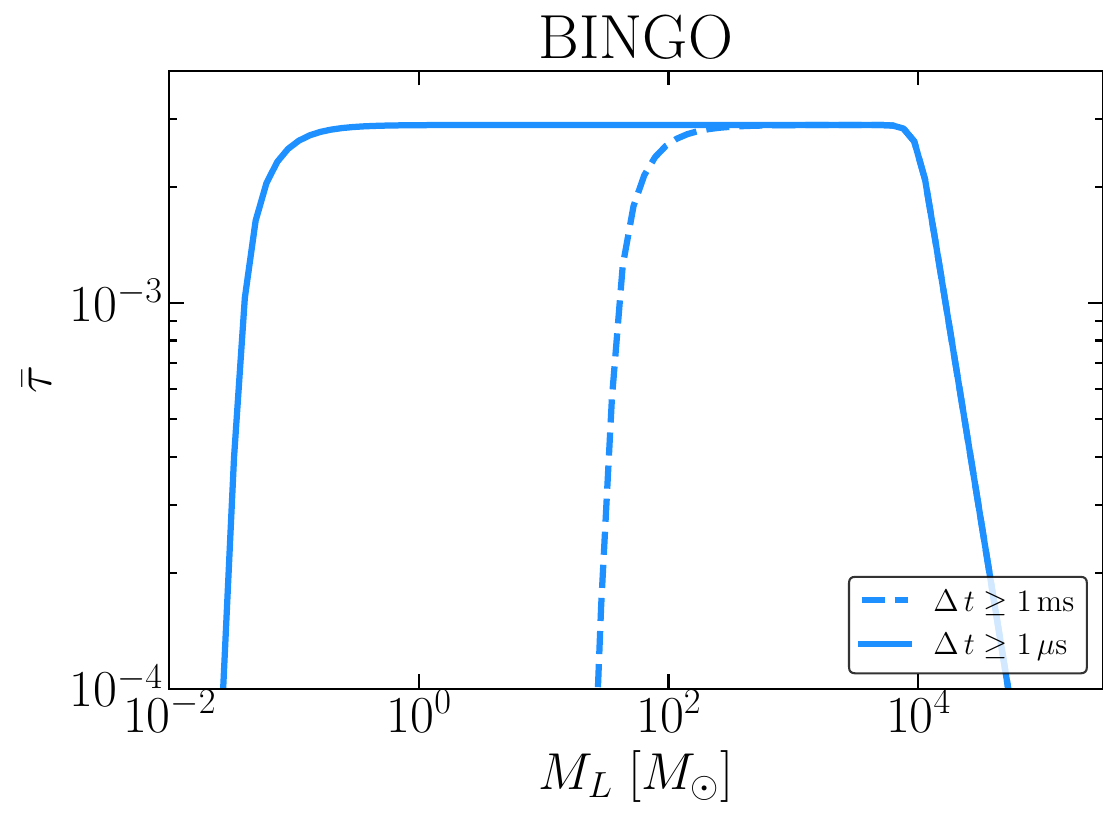}
    \caption{Total optical depth for LOFAR, FAST, and BINGO. In these graphics, we consider the best time resolution constraints for each radio telescope and $f_{\mathrm{PBH}} = 1$. The solid and dashed lines respectively correspond to the lowest and highest best time resolution in Tab.~\ref{tab:01}.} The asymptotic values are $\bar{\tau} \sim 0.0014$ for LOFAR, and $\bar{\tau} \sim 0.003$ for BINGO and FAST. 
    \label{fig:03}
\end{figure}

Now, given a distribution of FRBs, we can compute the integrated optical depths \eqref{eq:opticaldepthint}. We present their main features in Fig.~\ref{fig:03}. There, we observe that the integrated optical depths grow monotonically as the mass of the lenses increases, as found by Muñoz et al. \cite{Munoz_2016}. The values of the plateaus mean that we expect $\sim 6$ lensing events detected by LOFAR with masses starting at $M_{\rm PBH} \sim 1\,{M_{\odot}}$ for $N_{\mathrm{FRB}} = 4500$. For FAST and BINGO, we estimate $\sim 3$ lensing events detected for $N_{\mathrm{FRB}} = 900$, with PBH masses larger than $M_{\rm PBH} \sim 30 \,{M_{\odot}}$ and $M_{\rm PBH} \sim 0.2 \,{M_{\odot}}$, respectively. From Fig.~\ref{fig:03}, we also see the influence of time resolution and of the high-mass cutoff on the bounds of the mass ratios that can be detected by the three telescopes in this forecast. Namely, the higher the time resolution, the lower the mass of the lens that can be resolved.

\begin{figure}
    \centering
    \includegraphics[width=0.49\linewidth]{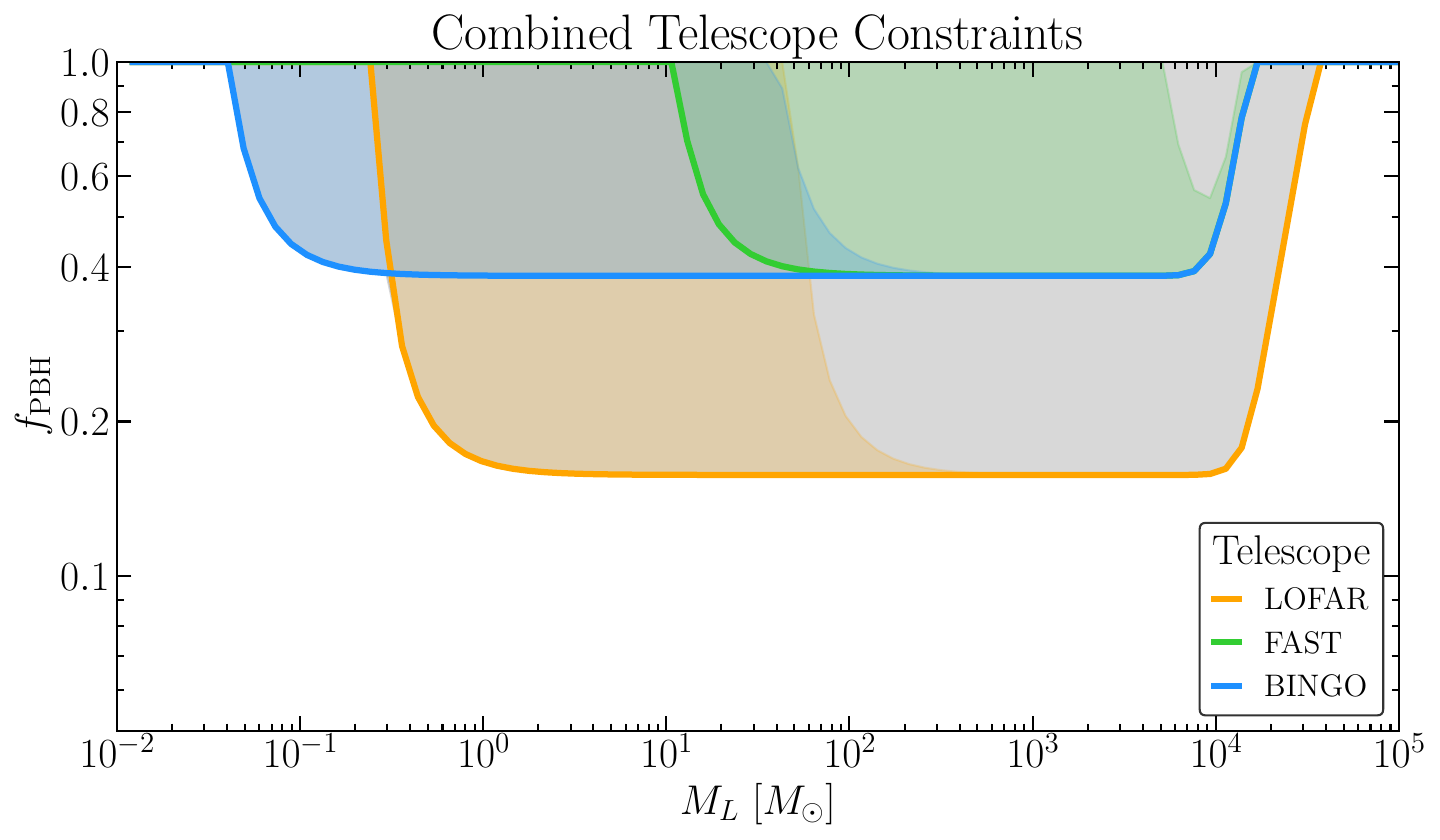}
    \includegraphics[width=0.49\linewidth]{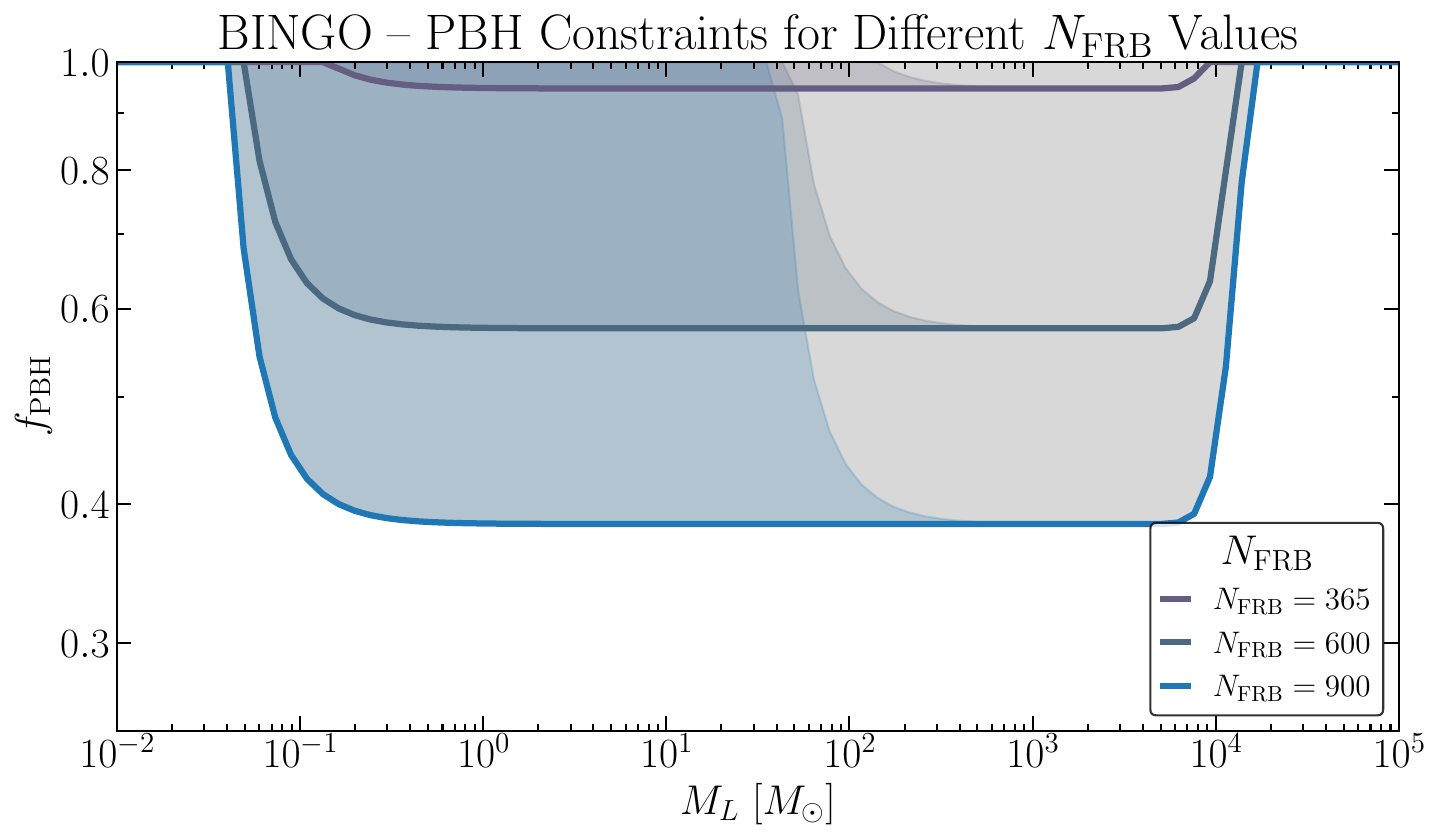}
    \caption{Forecast for the fraction of primordial black holes allowed as point lenses. \textbf{Left:} The colored shaded regions represent the best constraints of exclusion for LOFAR (orange), FAST (green), and BINGO (blue) using the lowest best time resolution of Tab.~\ref{tab:01}. The gray shaded regions correspond to the highest best time resolution of Tab.~\ref{tab:01}, instead. \textbf{Right:} Forecast for the fraction of primordial black holes for BINGO, considering $N_{\mathrm{FRB}} = 350$ (purple), $N_{\mathrm{FRB}} = 600$ (darker blue), and $N_{\mathrm{FRB}} = 900$ (lighter blue). The colored shade regions represent the best constraints for $f_{\rm PBH}$ considering a time resolution between $1\,\mu {\rm s} \,- \, 1\, {\rm ms}$.}
    \label{fig:04}
\end{figure}

To find the constraints on the fraction of PBHs considering a null-detection, we work in the optically thin regime approximation, where the probability of a FRB being lensed is $P_{\mathrm{lens}} = 1 - e^{-\bar{\tau}} \approx \bar{\tau}$. By following the prescription adopted by Mu\~noz et al. \cite{Munoz_2016}, and by Kalita et al. \cite{Kalita_2023}, we will initially constrain the maximum allowed PBH fraction related to this probability as
\begin{equation}\label{eq:fpbhmax}
    f_{\mathrm{PBH\,max}} (M_{\mathrm{PBH}}) = \left(N_{\mathrm{FRB}} \, \bar{\tau}(M_{\mathrm{PBH}})\right)^{\,-1}\,,
\end{equation}
where, from now on, we will set the lens mass to the PBH mass, namely $M_L=M_{\rm PBH}$. This maximum fraction is known as the one-event criterion and corresponds to the most optimistic scenario for our forecast. We also impose that $f_{\mathrm{PBH\,max}} \leq 1$ for consistency. We show the forecast for $f_{\mathrm{PBH\,max}}$ versus $M_{\rm PBH}$ in Fig. \ref{fig:04}. In the shaded regions of Fig.~\ref{fig:04}, we present the exclusion regions for $f_{\mathrm{PBH\,max}}$ for each radio telescope. The shaded colored regions unveil the lowest best interval of time resolution of LOFAR (orange), FAST (green), and BINGO (blue), to compute $f_{\mathrm{PBH}}$. The shaded gray regions mark the constraints use the highest best interval of time resolution of Tab.~\ref{tab:01}, also see Fig.~\ref{fig:03}.

From the left panel of Fig.~\ref{fig:04}, we see that the upgrades of LOFAR could constrain a maximum fraction of $f_{\rm PBH}<0.14$ for $M_{\rm PBH}\sim 1 - 10^{2}\,{M_{\odot}}$. For the next upgraded version of FAST operating together with the BINGO-ABDUS consortium, we would have the possibility of constraining a maximum fraction of $f_{\rm PBH}<0.39$ for $M_{\rm PBH}\sim 10 - 10^{4}\,{M_{\odot}}$. Note that, since the time resolution reported by FAST is $128\,{\rm ms}- 270\,\mu {\rm s}$, it can only provide constraints for lenses with high masses. In the case of BINGO-ABDUS, the collaboration aims to achieve a $\mathrm{SNR} = 10$ (in a more conservative scenario) and a time resolution of between $\mu{\rm s}$ and $\rm ms$. Therefore, BINGO-ABDUS would be able to constrain a maximum fraction of $f_{\rm PBH}<0.39$, the same as FAST, but with a higher time resolution, extending the constraint to lower masses and covering an optimistic range between $ 10^{-1}\,{M_{\odot}}$ and $10^{2}\,{M_{\odot}}$.

\begin{figure}
    \centering
    \includegraphics[width=0.49\linewidth]{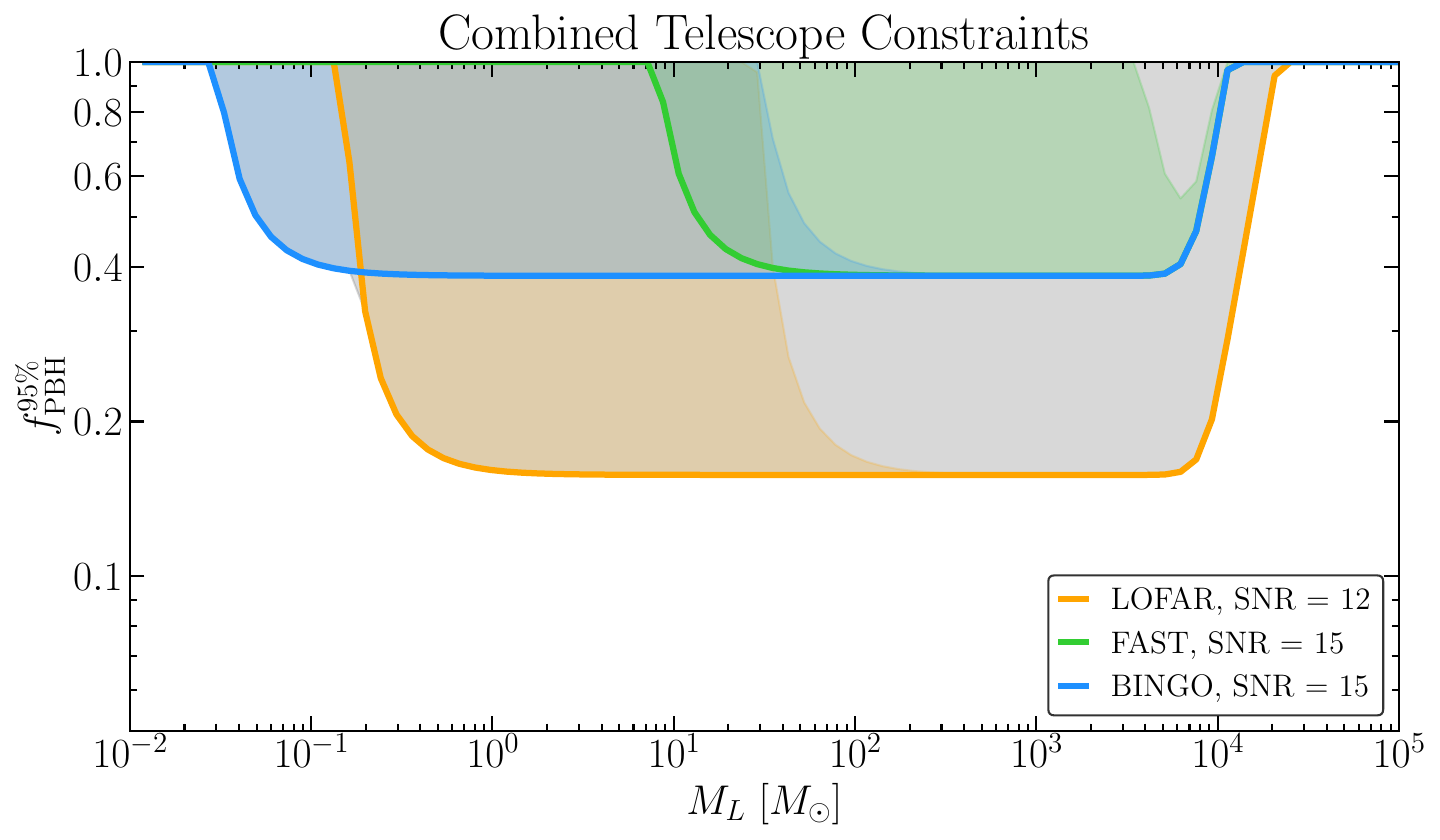}
    \caption{Forecast for the fraction of primordial black holes allowed as point lenses with $95\,\%$ confidence-level Poisson statistics. The colored shaded regions represent the best constraints of exclusion using the time resolution of Tab.~\ref{tab:01}, $\mathrm{SNR} = 12, \,15,\,15$, and $N_{\mathrm{FRB}} = 5000,\, 1500, \, 1500$ for LOFAR (orange), FAST (green), and BINGO (blue), respectively.}
    \label{fig:045}
\end{figure}

\begin{figure}
    \centering
    \includegraphics[width=0.69\linewidth]{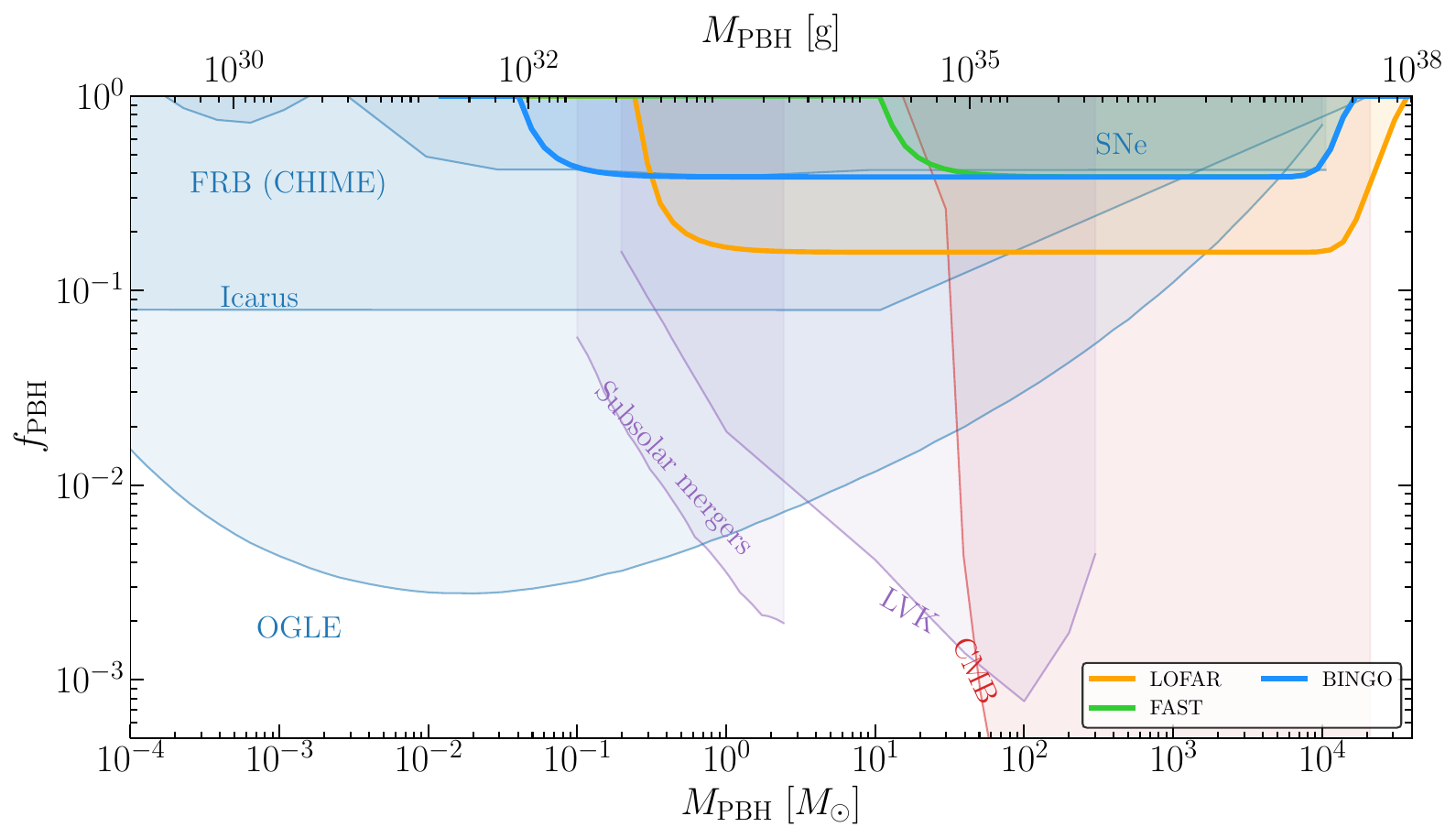}
    \caption{Constraints on the fraction of primordial black holes from different surveys combined with our forecast. The colored shaded regions represent the best constraints of exclusion provided by FRB lensing using CHIME \cite{Leung_2022} (dark blue), SN lensing \cite{Zumalacarregui:2017qqd} (dark blue), Icarus caustic-crossing \cite{Oguri:2017ock} (blue), OGLE \cite{Mroz:2024wag} (light blue), Subsolar mergers \cite{Nitz:2022ltl} (purple), LIGO/VIRGO/KAGRA mergers \cite{Kavanagh:2018ggo} (light purple), CMB \cite{Serpico:2020ehh} (red), see also \cite{Agius:2024ecw}. We used the public code \textsc{PBHbounds} \cite{PBHbounds,Green:2020jor} to plot existing constraints. The forecasts derived in this work are LOFAR (highlighted orange), FAST (highlighted green), and BINGO (highlighted blue). For the telescope constraints, we consider the lowest-best time resolution in Tab.~\ref{tab:01}.}
    \label{fig:05}
\end{figure}

The left panel of Fig.~\ref{fig:04} also shows that, in the large-mass regime, the constraints become independent of $M_{L}$, forming plateaus. To verify such a behavior, let us go back to Eq. \eqref{sec4_eq2}, where we can see that the product
\begin{equation}
    n_L\,\sigma = \frac{4\,\pi\,G}{c^2}\,\frac{D_L\,D_{LS}}{D_{S}}\,\rho_0\,f_{\mathrm{PBH}}\,\Omega_c\,\left(y_{\mathrm{max}}^2 (\mu)-y_{\mathrm{min}}^2(M_L,z_L)\right)\,,
\end{equation}
has a mass dependence only through the computation of $y_{\mathrm{min}}$. Therefore, by considering that at sufficiently large mass $y_{\mathrm{max}}\gg y_{\mathrm{min}}$, we find
\begin{equation}
    n_L\,\sigma \rightarrow \frac{3}{2}\frac{H_0^2}{c^2}\,\frac{D_L\,D_{LS}}{D_{S}}\,f_{\mathrm{PBH}}\,\Omega_c\,y_{\mathrm{max}}^2(\mathrm{SNR})\,,
\end{equation}
where $y_{\mathrm{max}}$ is a constant determined by the signal-to-noise ratio, as  we see in Eq. \eqref{s3_eq1_5}. By substituting the previous equation in the definition of the total optical depth, we obtain
\begin{equation}
    \bar{\tau} = \frac{3}{2}\frac{H_0^2}{c}\,f_{\mathrm{PBH}}\,\Omega_c\,y_{\mathrm{max}}^2(\mathrm{SNR})\,{\cal I}\,;\qquad {\cal I} = \int \bar{N}(z)dz\,\int_0^z\frac{D_LD_{LS}}{D_S}\frac{dz_L}{(1+z_L)H(z_L)}\,.
\end{equation}
Since the last integral is mass independent, ${\cal I}$ will be a constant; consequently, $\bar{\tau}$ and $f_{\mathrm{PBH}}$ are constants in this mass regime, resulting in the plateaus observed in Figs.~\ref{fig:04} - ~\ref{fig:05}. These plateaus and even the bounds of our forecast can be degraded by different reasons, such as scattering, plasma screens, decoherence in the intergalactic medium \cite{Leung_2022}, and finite search windows \cite{Oguri:2022fir}. Our approach combined the procedures introduced by \cite{Leung_2022, Kalita_2023, Oguri:2022fir} on the integration of the total optical depth, to build a search window based on the time resolution of the radio telescopes (for the left-hand side limit), and high-mass cutoff $M_{\mathrm{max}}$ (for the right-hand side limit).

To understand the influence of the suppressions on constraints and high-mass sensitivity, it is relevant to increase the number of detected FRBs and the resolution of current telescopes, thereby enabling a better interpretation of the intergalactic medium, the limitations of sensitivity due to the maximum time delay, and their real impact on FRB dispersion. By increasing the number of detected events, we will be able to corroborate the characterization techniques used to distinguish transient light curves induced by gravitational lensing of FRBs from complex temporal structures.

In the right panel of Fig.~\ref{fig:04}, we depict a forecast for BINGO varying the detected number of FRB events, namely $N_{\mathrm{FRB}} = 350$ (purple), $N_{\mathrm{FRB}} = 600$ (darker blue), and $N_{\mathrm{FRB}} = 900$ (lighter blue). We find that BINGO needs to detect more than $350$ FRBs to be a competitive radio telescope for constraining $f_{\rm PBH}$. As we see from Fig.~\ref{fig:04}, if $N_{\mathrm{FRB}} = 350$, we find that $f_{\mathrm{PBH}}<0.95$ for $M_{\rm PBH}> 1\,\mathrm{M_{\odot}}$.
For $N_{\mathrm{FRB}} = 600$, the bound goes down to $f_{\mathrm{PBH}} <0.58$ for $M_{\rm PBH}> 10^{-1}\,\mathrm{M_{\odot}}$. For $N_{\mathrm{FRB}} = 900$, which is approximately the maximum number of expected events, we recover the results of the left-hand side of Fig.~\ref{fig:04}. We see that, for BINGO's resolution, a larger number of FRBs impacts not only the bounds on $f_{\mathrm{PBH}}$, but also the range of $M_{\rm PBH}$. 

We can also extend the forecast to a more realistic scenario by considering conventional $95\,\%$ confidence-level Poisson statistics as the upper bound for the PBH fraction \cite{Oguri:2022fir}, whose equation is 
\begin{equation}\label{eq:fpbhmax_2}
    f_{\mathrm{PBH\,max}}^{95\%} (M_{\mathrm{PBH}}) = 3.05\,\times \left(N_{\mathrm{FRB}} \, \bar{\tau}(M_{\mathrm{PBH}})\right)^{\,-1}\,.
\end{equation}
The last equation means that our quoted limits depicted in Fig.~\ref{fig:04}  will be weaker by a factor of $3$. Therefore, in this case the BINGO and FAST bounds for $900$ FRBs will be negligible, and LOFAR will be able to constrain $f_{\mathrm{PBH}} < 0.43$. However, we can still provide a scenario similar to the most optimistic one by increasing the $\mathrm{SNR}$ of the three radio telescopes and the number of detected FRBs. This new analysis is shown in Fig.~\ref{fig:045}, where we used $\mathrm{SNR} = 12, \,15,\,15$, and $N_{\mathrm{FRB}} = 5000,\, 1500, \, 1500$ for LOFAR, FAST, and BINGO, respectively.  

To compare our forecast with current observational constraints on PBHs, we generated Fig. \ref{fig:05}. There we can observe bounds generated by FRB/CHIME \cite{Leung_2022}, SNe \cite{Zumalacarregui:2017qqd}, Icarus \cite{Oguri:2017ock}, OGLE \cite{Niikura:2017zjd, Mroz:2024mse}, Subsolar mergers \cite{Nitz:2022ltl}, LIGO/VIRGO/KAGRA \cite{Green:2020jor, Andres-Carcasona:2024wqk}, and CMB \cite{Serpico:2020ehh}. These bounds cover a broad range of PBHs' masses, here highlighted from $10^{-4}–10^{4}\,{M_{\odot}}$. They can also constrain fractions of PBHs $< 10^{-3}$, as we can see from the bounds provided by OGLE and CMB. By comparing our forecast with these bounds, we find it is compatible with forecasts for FRBs and Supernovae reported so far. The time resolution and mass scales presented here are in agreement with the forecast of the BURSTT (Bustling Universe Radio Survey Telescope in Taiwan) collaboration \cite{Ho:2023feo}. We also realize that the three radio telescopes can contribute to different surveys in a new and independent way. Moreover, the precision of this forecast can be enhanced as new FRBs are reported and characterized in the near future with better instrumental resolution.

\section{Conclusions and discussion}\label{final_remarks}

In the next years, the radio telescopes LOFAR, FAST, and BINGO are expected to detect of the order of $1000$ FRB events (see Tab.~\ref{tab:01}). A null-detection of FRB lensed events will constrain the fraction of PBHs as dark matter, $f_{\rm PBH}\lesssim {\cal O}(0.1)$. More concretely, we showed that LOFAR2.0 is expected to yield $f_{\rm PBH}<0.16$ for $M_{\rm PBH}>1\, M_\odot$, while FAST Core Array together with BINGO (in the BINGO-ABDUS joint phase \cite{bingoabdus-2023}) should constrain $f_{\rm PBH}<0.39$ for $M_{\rm PBH}>0.1\, M_\odot$. Even considering an exclusion scenario with $2\sigma$ confidence level pointed out in Eq. \eqref{eq:fpbhmax_2}, we can reproduce similar constraints for $f_{\mathrm{PBH}}^{95\%} $ by increasing the $\mathrm{SNR}$ and the $N_{\mathrm{FRB}}$, as presented in Fig.\ref{fig:045}. While these forecasts are not yet competitive with current constraints--in our same PBH mass range microlensing of stars yields $f_{\rm PBH}<0.01$ for $0.01\, M_\odot<M_{\rm PBH}<100 M_\odot$, and CMB bounds $f_{\rm PBH}<10^{-8}$ for $M_{\rm PBH}>100 M_\odot$ \cite{Carr:2020gox,Profumo_2025,Carr:2026hot}--, FRB lensing is a new and independent way to test the PBH scenario, which will improve in the future. For instance, we find that reaching $N_{\rm FRB}=2\times 10^{4}$ and $N_{\rm  FRB}=2\times 10^{5}$ would yield $f_{\rm PBH}<10^{-2}$ and $f_{\rm PBH}<10^{-3}$ respectively, in agreement with Ref.~\cite{Munoz_2016}. We show our forecasts together with existing constraints in Fig.~\ref{fig:05}.

Note that, although our results build upon the work of Ref.~\cite{Munoz_2016}, the new aspect of our work is the use of concrete design features of these upcoming radio telescopes. We hope that our forecast may be useful for the operation of LOFAR, FAST, and BINGO in the near future, shaping possible interesting applications for their upcoming FRB data. Furthermore, the methodology we adopted in our work can be applied to other surveys \cite{Leung_2022} to include other types of gravitational lenses, subnanosecond detections \cite{Xiao:2024qay}, dispersion effects such as plasma screens \cite{Leung_2022, Kalita_2023}, and test gravitational theories \cite{Kalita_2023}. 

Lastly, in deriving our forecasts, we have assumed a monochromatic PBH mass function for simplicity. Considering a broad mass function will not qualitatively change the constraints on the magnitude of $f_{\rm PBH}$ but may change the PBH mass range probed depending on the mass function width. Furthermore, we have not considered possible decoherence effects from the interaction of FRBs with the intervening intergalactic medium. These effects mainly deteriorate the constraining power of FRB lensing in the large mass lens regime \cite{Leung_2022}. Our forecasts of Fig.~\ref{fig:04} can be understood as the best case scenario for the radio telescopes considered. We leave a more detailed analysis for future work when new data becomes available.

{\acknowledgments}
JRLS acknowledges support from CNPq (Grants 309494/2021-4 and 302190/2025-2), FAPESQ-PB (Grant 1356/2024), CAPES Finance Code 001, and Alexander von Humboldt-Stiftung Foundation. JRLS also acknowledges the hospitality of the ITP at Leibniz University Hannover, and the ITP and ITA at the University of Heidelberg, where this work was built. A.R.Q. acknowledges FAPESQ-PB, and the support by CNPq under process number 310533/2022-8. G.D acknowledges support from the DFG under the Emmy-Noether program (project number 496592360) and the JSPS KAKENHI grant No. JP24K00624. The authors would like to thank the anonymous referee for the criticism and contributions, which enhanced the quality of this work.\\

\appendix

\section{Fast radio bursts dataset} \label{append_a}

In this appendix we list the FRBs used in our forecast. We include FRB names, redshift, DM and reference of origin in Tabs.~\ref{tab1} and \ref{tab2}.
\begin{table}[htbp!]
\hspace*{-1cm}     
    \begin{subtable}{.49\textwidth}
        \begin{tabular}{cccccc} \hline\hline
FRB & Redshift & $\rm DM_{obs}$ & $\rm DM_{MW}$ & Ref.\\
(name) & $z$ &  ($\rm pc\ cm^{-3}$) & ($\rm pc\ cm^{-3}$) & \\ \hline
20250316A & 0.0067 & 161.82 & 70 & \cite{Chime:2025arx}\\[-0.5mm] 
20171020A & 0.00867 & 114.1 & 38 & \cite{Mahony:2018ddp}\\[-0.5mm]
20220319D & 0.011228 & 110.98 & 133.3 & \cite{Law:2023ibd}\\[-0.5mm]
20231229A & 0.0190 & 198.5 & 58.2 & \cite{CHIMEFRB:2025ggb, Feng:2025arx}\\[-0.5mm] 
20240210A & 0.0237 & 283.75 & 31 & \cite{Shannon:2024pbu}\\[-0.5mm] 
20181220A & 0.0275 & 209.4 & 126 & \cite{Bhardwaj:2023vha,CHIMEFRB:2021srp}\\[-0.5mm] 
20231230A & 0.0298 & 131.4 & 61.58 & \cite{CHIMEFRB:2025ggb, Rui-Nan:2025TAJ}\\[-0.5mm]
20200120E & 0.03 & 87.82 & 30 & \cite{Bhardwaj:2021xaa}\\[-0.5mm]
20181030A & 0.03 & 103.5 & 30 & \cite{Bhardwaj:2021hgc,CHIMEFRB:2021srp}\\[-0.5mm]
20181223C & 0.03024 & 112.5 & 20 & \cite{Bhardwaj:2023vha}\\[-0.5mm] 
20190425A & 0.03122 & 128.2 & 49 & \cite{Bhardwaj:2023vha,CHIMEFRB:2021srp}\\[-0.5mm] 
20180916B & 0.0337 & 348.76 & 200 & \cite{Gordon:2023cgw,CHIMEFRB:2021srp}\\[-0.5mm] 
20230718A & 0.035 & 477 & 393 & \cite{glowacki2024h}\\[-0.5mm] 
20240201A & 0.0427 & 374.5 & 38 & \cite{Shannon:2024pbu}\\[-0.5mm] 
20220207C & 0.0430 & 262.38 & 79.3 & \cite{Law:2023ibd}\\[-0.5mm] 
20211127I & 0.0469 & 234.83 & 42 & \cite{Gordon:2023cgw}\\[-0.5mm] 
20201123A & 0.0507 & 433.55 & 251.93 & \cite{Rajwade:2022zkj}\\[-0.5mm] 
20230926A & 0.0553 & 222.8 & 52.62 & \cite{CHIMEFRB:2025ggb}\\[-0.5mm] 
20200223B & 0.06024 & 201.8 & 45.6 & \cite{Ibik:2023ugl,CHIMEFRB:2021srp}\\[-0.5mm] 
20190303A & 0.064 & 222.4 & 26 & \cite{Michilli:2022bbs,CHIMEFRB:2021srp}\\[-0.5mm] 
20231204A & 0.0644 & 221.0 & 29.73 & \cite{CHIMEFRB:2025ggb}\\[-0.5mm] 
20231206A & 0.0659 & 457.75 & 90.46 & \cite{CHIMEFRB:2025ggb, Chime:202509}\\[-0.5mm] 
20210405I & 0.066 & 565.17 & 516.1 & \cite{Driessen:2023lxj}\\[-0.5mm] 
20180814 & 0.068 & 189.4 & 87 & \cite{Michilli:2022bbs,CHIMEFRB:2021srp}\\[-0.5mm] 
20231120A & 0.07 & 438.9 & 43.8 & \cite{Sharma:2024fsq,Connor:2024mjg}\\[-0.5mm] 
20231005A & 0.0713 & 189.4 & 33.37 & \cite{CHIMEFRB:2025ggb}\\[-0.5mm] 
20190418A & 0.07132 & 184.5 & 71 & \cite{Bhardwaj:2023vha,CHIMEFRB:2021srp}\\[-0.5mm] 
20211212A & 0.0715 & 109 & 42 & \cite{Gordon:2023cgw,leung2025:11}\\[-0.5mm] 
20231123A & 0.0729 & 302.1 & 90 & \cite{CHIMEFRB:2025ggb, leung2025:11}\\[-0.5mm] 
20220912A & 0.0771 & 219.46 & 125 & \cite{DeepSynopticArrayTeam:2022rbq,Zhang:2023eui}\\[-0.5mm] 
20231011A & 0.0783 & 186.3 & 70 & \cite{CHIMEFRB:2025ggb, leung2025:11}\\[-0.5mm] 
20220509G & 0.0894 & 269.53 & 55.2 & \cite{Law:2023ibd}\\[-0.5mm] 
20230124 & 0.0940 & 590.6 & 38.5 & \cite{Connor:2024mjg}\\[-0.5mm] 
\hline\hline
\end{tabular}
    \end{subtable}%
    \hspace*{1cm}
    \begin{subtable}{.49\textwidth}
        \begin{tabular}{ccccc} \hline\hline
FRB & Redshift & $\rm DM_{obs}$ & $\rm DM_{MW}$ & Ref.\\
(name) & $z$ &  ($\rm pc\ cm^{-3}$) & ($\rm pc\ cm^{-3}$) & \\ \hline
20241228A & 0.1614 & 246.53 & 23.8 & \cite{Curtin:2025arx}\\[-0.5mm] 
20210603A & 0.177 & 500.15 & 40 & \cite{Cassanelli:2023hvg}\\[-0.5mm] 
20220529A & 0.1839 & 246.0 & 39.93 & \cite{gao2024measuring, li2025}\\[-0.5mm] 
20230311A & 0.1918 & 364.3 & 67.24 & \cite{CHIMEFRB:2025ggb, JiGuozhang2025}\\[-0.5mm] 
20220725A & 0.1926 & 290.4 & 31 & \cite{Shannon:2024pbu}\\[-0.5mm] 
20121102A & 0.19273 & 551.92 & 200 & \cite{Gordon:2023cgw, snelders2025}\\[-0.5mm] 
20221106A & 0.2044 & 343.8 & 35 & \cite{Shannon:2024pbu}\\[-0.5mm] 
20240215A & 0.21 & 549.5 & 48.0 & \cite{Connor:2024mjg}\\[-0.5mm] 
20230730A & 0.2115 & 312.5 & 85.16 & \cite{CHIMEFRB:2025ggb}\\[-0.5mm] 
20210117A & 0.214 & 729.0 & 34.0 & \cite{Bhandari:2022ton}\\[-0.5mm] 
20221027A & 0.229 & 452.5 & 47.2 & \cite{Connor:2024mjg}\\[-0.5mm] 
20191001A & 0.234 & 506.92 & 44.0 & \cite{Gordon:2023cgw,Bhandari:2020cde}\\[-0.5mm] 
20190714A & 0.2365 & 504.13 & 38 & \cite{Gordon:2023cgw, hussaini2025, wei2023}\\[-0.5mm] 
20221101B & 0.2395 & 490.7 & 131.2 & \cite{DeepSynopticArrayTeam:2023iev,Sharma:2024fsq,Connor:2024mjg}\\[-0.5mm] 
20220825A & 0.2414 & 651.24 & 79.7 & \cite{Law:2023ibd}\\[-0.5mm] 
20190520B & 0.2418 & 1204.7 & 60.2 & \cite{Gordon:2023cgw, Koch_Ocker_2022}\\[-0.5mm] 
20191228A & 0.2432 & 297.5 & 33 & \cite{Bhandari:2021pvj}\\[-0.5mm] 
20231017A & 0.2450 & 344.2 & 64.55 & \cite{CHIMEFRB:2025ggb}\\[-0.5mm] 
20220307B & 0.2481 & 499.15 & 128.25 & \cite{Law:2023ibd, Zhuge:2025arx}\\[-0.5mm] 
20221113A & 0.2505 & 411.4 & 91.7 & \cite{DeepSynopticArrayTeam:2023iev,Sharma:2024fsq,Connor:2024mjg}\\[-0.5mm] 
20220831A & 0.262 & 1146.25 & 126.7 & \cite{Connor:2024mjg}\\[-0.5mm] 
20231123B & 0.2625 & 396.7 & 40.2 & \cite{DeepSynopticArrayTeam:2023iev,Sharma:2024fsq,Connor:2024mjg}\\[-0.5mm] 
20230307A & 0.2710 & 608.9 & 37.6 & \cite{DeepSynopticArrayTeam:2023iev,Sharma:2024fsq,Connor:2024mjg}\\[-0.5mm] 
20221116A & 0.2764 & 640.6 & 132.3 & \cite{Sharma:2024fsq,Connor:2024mjg}\\[-0.5mm] 
20220105A & 0.2785 & 580 & 21.9 & \cite{Gordon:2023cgw, Connor:2024mjg}\\[-0.5mm] 
20210320C & 0.2797 & 384.59 & 39.2 & \cite{Gordon:2023cgw, Connor:2024mjg}\\[-0.5mm] 
20221012A & 0.2840 & 442.20 & 54.4 & \cite{Law:2023ibd, Connor:2024mjg}\\[-0.5mm] 
20240229A & 0.287 & 491.15 & 37.9 & \cite{Connor:2024mjg}\\[-0.5mm] 
20190102C & 0.2912 & 364.55 & 57.4 & \cite{Gordon:2023cgw, Connor:2024mjg}\\[-0.5mm] 
20220506D & 0.3005 & 396.93 & 84.5 & \cite{Law:2023ibd, Connor:2024mjg}\\[-0.5mm] 
20230501A & 0.3010 & 532.5 & 125.6 & \cite{DeepSynopticArrayTeam:2023iev,Connor:2024mjg}\\[-0.5mm] 
20230503E & 0.32 & 483.74 & 88 & \cite{Marazuela:2025arx}\\[-0.5mm] 
20180924B & 0.3214 & 361.42 & 40.5 & \cite{Gordon:2023cgw, wei2023}\\[-0.5mm] 
\hline\hline
\end{tabular}
    \end{subtable} 
 \caption{First part of the dataset of 131 localized FRBs reported by different radio telescopes.}
 \label{tab1}
\end{table}

\begin{table}
\hspace*{-1cm}     
    \begin{subtable}{.49\textwidth}
        \begin{tabular}{cccccc} \hline\hline
FRB & Redshift & $\rm DM_{obs}$ & $\rm DM_{MW}$ & Ref.\\
(name) & $z$ &  ($\rm pc\ cm^{-3}$) & ($\rm pc\ cm^{-3}$) & \\ \hline
20240310A & 0.127 & 601.8 & 36 & \cite{Shannon:2024pbu}\\[-0.5mm] 
20210807D & 0.1293 & 251.9 & 121 & \cite{Gordon:2023cgw, Shannon:2024pbu}\\[-0.5mm] 
20240114A & 0.13 & 527.7 & 49.7 & \cite{Kumar:2024svu}\\[-0.5mm] 
20240209A & 0.1384 & 176.49 & 55.5 & \cite{Shah:2024ywp}\\[-0.5mm] 
20210410D & 0.1415 & 578.78 & 56.2 & \cite{Gordon:2023cgw,Caleb:2023atr}\\[-0.5mm] 
20230203A & 0.1464 & 420.1 & 36.29 & \cite{CHIMEFRB:2025ggb}\\[-0.5mm] 
20231226A & 0.1569 & 329.9 & 145 & \cite{Shannon:2024pbu}\\[-0.5mm] 
20230526A & 0.157 & 316.4 & 50 & \cite{Shannon:2024pbu}\\[-0.5mm] 
20220920A & 0.158 & 314.99 & 40.3 & \cite{Law:2023ibd}\\[-0.5mm] 
20200430A & 0.1608 & 380.25 & 27 & \cite{Hiramatsu:2022tyn, wei2023}\\[-0.5mm]
20230703A & 0.1184 & 290.74 & 57.44 & \cite{CHIMEFRB:2025ggb, Chime:202509}\\[-0.5mm] 
20240213A & 0.1185 & 357.4 & 40.1 & \cite{Connor:2024mjg}\\[-0.5mm] 
20240318A & 0.12 & 256.4 & 37 & \cite{Shannon:2024pbu}\\[-0.5mm] 
20230222A & 0.1223 & 706.1 & 134.2 & \cite{CHIMEFRB:2025ggb, Zhuge:2025arx}\\[-0.5mm] 
20190110C & 0.1224 & 221.6 & 37.1 & \cite{Ibik:2023ugl,CHIMEFRB:2021srp}\\[-0.5mm] 
20230628A & 0.1265 & 345.15 & 39.1 & \cite{DeepSynopticArrayTeam:2023iev,Sharma:2024fsq,Connor:2024mjg}\\[-0.5mm]
20190711A & 0.5217 & 587.9 & 56.4 & \cite{Gordon:2023cgw,Connor:2024mjg}\\[-0.5mm] 
20230216A & 0.5310 & 828.0 & 38.5 & \cite{Sharma:2024fsq,Connor:2024mjg}\\[-0.5mm] 
20230814B & 0.5535 & 696.4 & 104.9 & \cite{Connor:2024mjg} \\[-0.5mm] 
20221219A & 0.5540 & 706.7 & 44.4 & \cite{DeepSynopticArrayTeam:2023iev,Sharma:2024fsq,Connor:2024mjg}\\[-0.5mm] 
20190614D & 0.60 & 959.2 & 83.5 & \cite{Law:2020cnm,Hiramatsu:2022tyn}\\[-0.5mm] 
20231010A & 0.61 & 442.59 & 41 & \cite{Marazuela:2025arx}\\[-0.5mm] 
20220418A & 0.6220 & 623.45 & 36.7 & \cite{Law:2023ibd, Connor:2024mjg}\\[-0.5mm] 
20220224C & 0.6271 & 1140.2 & 52 & \cite{Marazuela:2025arx}\\[-0.5mm]
20231223C & 0.1059 & 165.8 & 48 & \cite{CHIMEFRB:2025ggb, leung2025:11}\\[-0.5mm] 
20201124A & 0.098 & 413 & 123 & \cite{Gordon:2023cgw,Lanman:2021yba}\\[-0.5mm] 
20230708A & 0.105 & 411.51 & 50 & \cite{Shannon:2024pbu}\\[-0.5mm]  
20191106C & 0.10775 & 333.2 & 25 & \cite{Ibik:2023ugl}\\[-0.5mm] 
20231128A & 0.1079 & 331.6 & 25 & \cite{CHIMEFRB:2025ggb, leung2025:11}\\[-0.5mm] 
20230222B & 0.11 & 187.8 & 28 & \cite{CHIMEFRB:2025ggb, leung2025:11}\\[-0.5mm] 
20231201A & 0.1119 & 169.4 & 70 & \cite{CHIMEFRB:2025ggb, leung2025:11}\\[-0.5mm] 
20220914A & 0.1139 & 631.28 & 55.2 & \cite{Law:2023ibd}\\[-0.5mm] 
20190608B & 0.1178 & 338.7 & 33 & \cite{Gordon:2023cgw,Hiramatsu:2022tyn, Chittidi_2021}\\[-0.5mm]
\hline\hline
\end{tabular}
    \end{subtable}%
    \hspace*{1cm}
    \begin{subtable}{.49\textwidth}
        \begin{tabular}{ccccc} \hline\hline
FRB & Redshift & $\rm DM_{obs}$ & $\rm DM_{MW}$ & Ref.\\
(name) & $z$ &  ($\rm pc\ cm^{-3}$) & ($\rm pc\ cm^{-3}$) & \\ \hline
20230613A & 0.3923 & 483.51 & 30 & \cite{Marazuela:2025arx}\\[-0.5mm] 
20220204A & 0.4 & 612.2 & 50.7 & \cite{DeepSynopticArrayTeam:2023iev,Sharma:2024fsq,Connor:2024mjg} \\[-0.5mm] 
20240208A & 0.4 & 260.2 & 98 & \cite{Shannon:2024pbu}\\[-0.5mm] 
20230712A & 0.4525 & 586.96 & 39.2 & \cite{DeepSynopticArrayTeam:2023iev,Sharma:2024fsq,Connor:2024mjg}\\[-0.5mm] 
20230907D & 0.4638 & 1030.79 & 29 & \cite{Marazuela:2025arx}\\[-0.5mm] 
20181112A & 0.4755 & 589.27 & 41.7 & \cite{Gordon:2023cgw, Connor:2024mjg}\\[-0.5mm] 
20231020B & 0.4775 & 952.2 & 34 & \cite{Marazuela:2025arx}\\[-0.5mm] 
20220310F & 0.479 & 462.15 & 46.3 & \cite{Law:2023ibd, Connor:2024mjg} \\[-0.5mm] 
20220918A & 0.491 & 656.8 & 41 & \cite{Shannon:2024pbu}\\[-0.5mm] 
20231210F & 0.5 & 720.6 & 32 & \cite{Marazuela:2025arx}\\[-0.5mm]
20230902A & 0.3619 & 440.1 & 34 & \cite{Shannon:2024pbu}\\[-0.5mm] 
20200906A & 0.3688 & 577.84 & 35.8 & \cite{Gordon:2023cgw,Hiramatsu:2022tyn}\\[-0.5mm] 
20240119A & 0.37 & 483.1 & 37.9 & \cite{Connor:2024mjg}\\[-0.5mm] 
20220330D & 0.3714 & 468.1 & 38.6 & \cite{Sharma:2024fsq,Connor:2024mjg}\\[-0.5mm] 
20190611B & 0.3778 & 322.2 & 57 & \cite{Gordon:2023cgw, Shannon:2024pbu}\\[-0.5mm] 
20220501C & 0.381 & 449.5 & 31 & \cite{Shannon:2024pbu}\\[-0.5mm]
20190523A & 0.6600 & 760.8 & 37.2 & \cite{Ravi:2019alc, Connor:2024mjg}\\[-0.5mm] 
20220222C & 0.853 & 1071.2 & 56 & \cite{Marazuela:2025arx}\\[-0.5mm] 
20240123A & 0.968 & 1462.0 & 90.3 & \cite{Connor:2024mjg}\\[-0.5mm] 
20221029A & 0.9750 & 1391.05 & 43.9 & \cite{DeepSynopticArrayTeam:2023iev,Sharma:2024fsq,Connor:2024mjg} \\[-0.5mm] 
20220610A & 1.015 & 1458.1 & 30.9 & \cite{Ryder:2022qpg, Connor:2024mjg}\\[-0.5mm] 
20230521B & 1.354 & 1342.9 & 138.8 & \cite{Shannon:2024pbu,Connor:2024mjg}\\[-0.5mm] 
20240304B & 2.148 & 2458.2 & 28.1 & \cite{Caleb:2025arx}\\[-0.5mm]
20230626A & 0.3270 & 451.2 & 39.2 & \cite{DeepSynopticArrayTeam:2023iev,Sharma:2024fsq,Connor:2024mjg}\\[-0.5mm] 
20231025B & 0.3238 & 368.7 & 48.59 & \cite{CHIMEFRB:2025ggb, Zhuge:2025arx}\\[-0.5mm] 
20230125D & 0.3265 & 640.08 & 88 & \cite{Marazuela:2025arx}\\[-0.5mm] 
 
20180301A & 0.3304 & 536 & 151.6 & \cite{Gordon:2023cgw,Price:2019fmc, Connor:2024mjg}\\[-0.5mm] 
20231220A & 0.3355 & 491.2 & 49.9 & \cite{Connor:2024mjg}\\[-0.5mm] 
20211203C & 0.3439 & 635.0 & 63.6 & \cite{Gordon:2023cgw, Connor:2024mjg}\\[-0.5mm] 
20220208A & 0.3510 & 437.0 & 101.6 & \cite{Sharma:2024fsq,Connor:2024mjg}\\[-0.5mm] 
20220726A & 0.3610 & 686.55 & 89.5 & \cite{DeepSynopticArrayTeam:2023iev,Sharma:2024fsq,Connor:2024mjg}\\[-0.5mm] 
20220717A & 0.3630 & 637.34 & 118.33 & \cite{Rajwade:2024ozu, Xu_2025, Zhuge:2025arx}\\[-0.5mm]
 & &  &  & \\[-0.5mm]
\hline\hline
\end{tabular}
    \end{subtable} 
 \caption{Second part of the dataset of 131 localized FRBs reported by different radio telescopes.}
 \label{tab2}
\end{table}
\newpage
\bibliography{bibli}

\end{document}